\pdfoutput=0
\documentclass[sigconf]{acmart}
\renewcommand\footnotetextcopyrightpermission[1]{} 
\AtBeginDocument{%
  \providecommand\BibTeX{{%
    \normalfont B\kern-0.5em{\scshape i\kern-0.25em b}\kern-0.8em\TeX}}}

\usepackage{algorithmic}
\usepackage[linesnumbered,ruled]{algorithm2e}
\usepackage{subfig}
\usepackage{graphics}
\usepackage{graphicx}
\usepackage{amsmath}
\newcommand{\signed}{\texttt{SIGNED}~}
\usepackage{todonotes}
\usepackage{multirow}
\usepackage{subfloat}

\begin{document}

\title{SIGNED: A Challenge-Response Based Interrogation Scheme for Simultaneous Watermarking and Trojan Detection}

\author{Abhishek Nair$^{*}$, Patanjali SLPSK$^{+}$, Chester Rebeiro$^{*}$, Swarup Bhunia$^{+}$}
\email{abhisheknair1729@gmail.com,patanjal.sristil@ufl.edu,chester@cse.iitm.ac.in,swarup@ece.ufl.edu}
\affiliation{
\institution{$^{+}$University of Florida}
\institution{$^{*}$Indian Institute of Technology,Madras}
}

\renewcommand{\shortauthors}{Nair et al.}

\begin{abstract}
 The emergence of distributed manufacturing ecosystems for electronic hardware involving untrusted parties has given rise to diverse trust issues. In particular, IP piracy, overproduction, and hardware Trojan attacks pose significant threats to digital design manufacturers. Watermarking has been one of the solutions employed by the semiconductor industry to overcome many of the trust issues. However, current watermarking techniques have low coverage, incur hardware overheads, and are vulnerable to removal or tampering attacks. Additionally, these watermarks cannot detect Trojan implantation attacks where an adversary alters a design for malicious purposes. We address these issues in our framework called \signed: Secure Lightweight Watermarking Scheme for Digital Designs. 
 \signed~relies on a challenge-response protocol based interrogation scheme for generating the watermark. \signed  identifies sensitive regions in the target netlist and samples them to form a compact signature that is representative of the functional and structural characteristics of a design. We show that this signature can be used to simultaneously verify, in a robust manner, the provenance of a design, as well as any malicious alterations to it at any stage during design process. We evaluate~\signed on the ISCAS85 and ITC benchmark circuits and obtain a detection accuracy of 87.61\% even for modifications as low as 5-gates. We further demonstrate that \signed can benefit from integration with a logic locking solution, where it can achieve increased protection against removal/tempering attacks and incurs lower overhead through judicious reuse of the locking logic for watermark creation. 
\end{abstract}


\maketitle

\section{Introduction}
\begin{figure*}[ht!]%
    \centering
    \subfloat[\centering Watermark Insertion]{{\includegraphics[width=.45\linewidth]{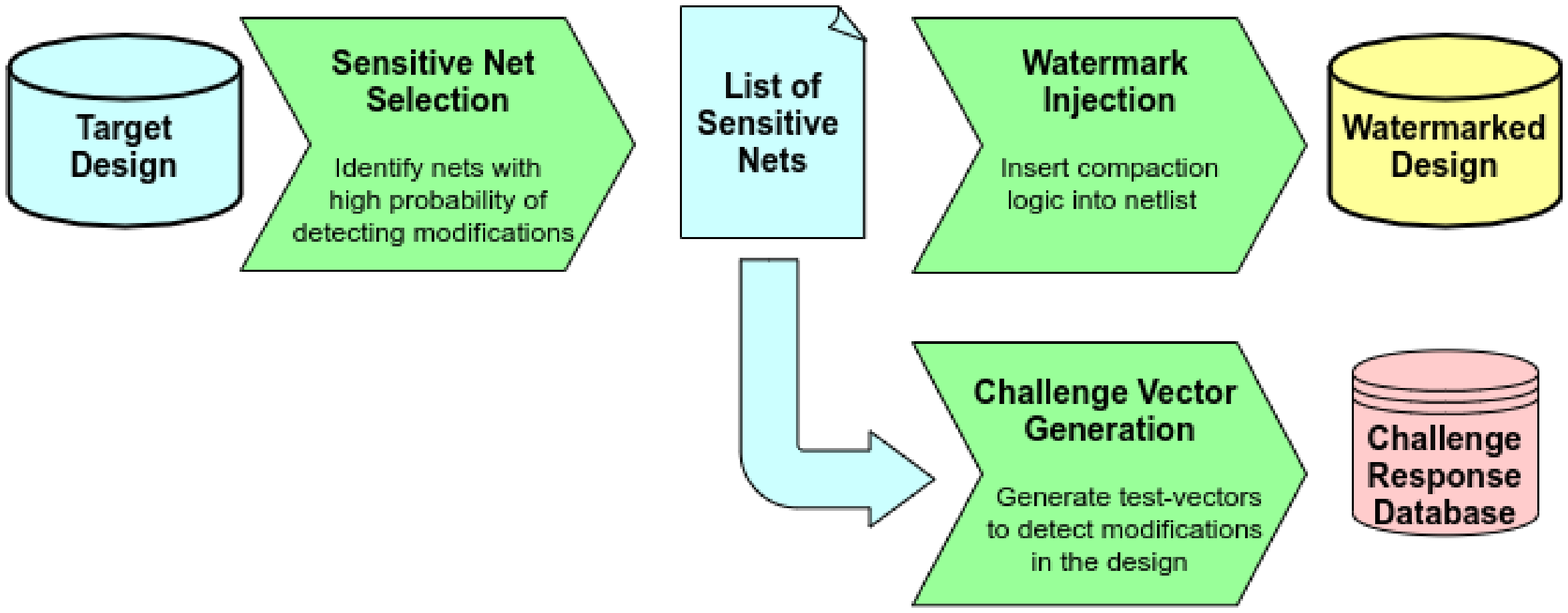} }}%
    \qquad
    \subfloat[\centering Watermark Authentication]{{\includegraphics[width=.34\linewidth]{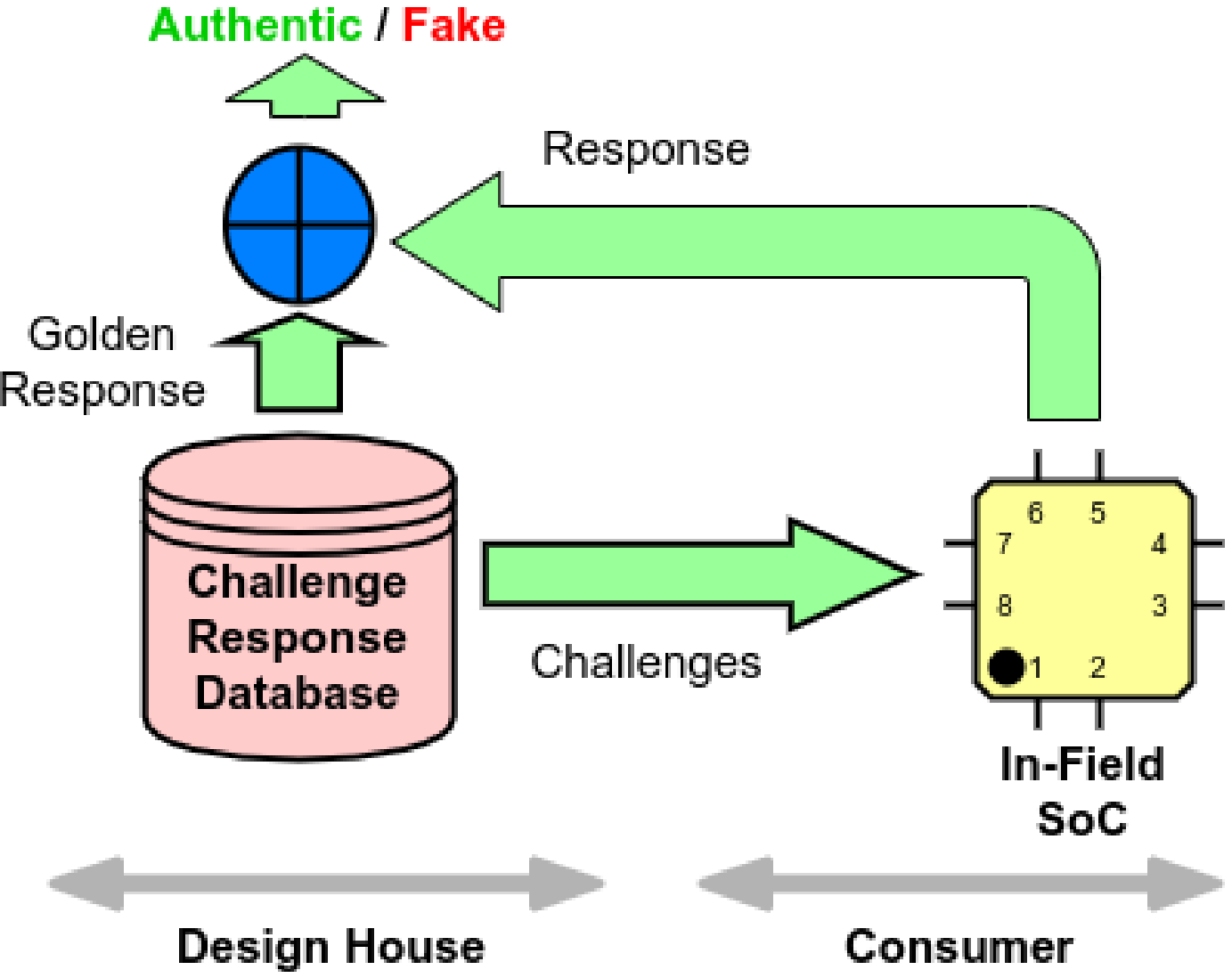} }}%
    \caption{The \signed~flow has two phases: A. watermark insertion and B. design authentication. The insertion is done by the design house. The authentication is done by the consumer with the help of the design house.}%
    \label{fig:overallflow}%
\end{figure*}
\label{sec:intro}
\noindent The increasing complexity of modern Intellectual Property blocks has resulted in a globally distributed manufacturing ecosystem. Although this entails considerable economic benefits, it increases the risk of an untrusted intermediary gaining access to the circuit. This can potentially result in IP piracy, counterfeiting and overproduction. Another security concern is that the untrusted entity may be able to insert a hardware Trojan into the design - either by tampering with the foundry mask or through a third-party IP that is added to the design.

\noindent IP Watermarking is a promising solution employed by the EDA industry to overcome the challenges of IP piracy, counterfeiting and overproduction. The watermarking process involves the addition of a unique signature to the original design which would enable the IP vendor to verify the provenance of the device at any stage in the supply-chain. The unique signature is usually added by carefully modifying a specific property of the original design. For example, recent works \cite{survey1}\cite{survey2}
have focused on either modifying the Finite State Machine (FSM) of the design or adding additional constraints during the EDA flow. However, a significant drawback of these techniques is the limited coverage offered by the watermarking scheme. That is, the watermarking techniques are unable to detect whether a part of the IP has been modified. Consequently, an adversary could tamper with the circuit by removing critical functionality or by implanting hardware Trojans. Trojans are malicious circuits introduced by an adversary into the original design. Existing watermarking schemes either fail to detect these Trojans (FSM-based techniques where the Trojan does not interfere with the control logic) or are modified by the Trojans and therefore cannot prove the authenticity of the circuit (Since now there is no way to distinguish between a modified circuit and a completely different circuit).

\noindent In order to address this issue, we propose \signed a \textbf{S}ecure l\textbf{IG}htweight watermarki\textbf{N}g schem\textbf{E} for \textbf{D}igital Designs. \signed is a hardware watermarking scheme that can detect hardware Trojans and malicious modifications while introducing low overheads. \signed can be utilized at any stage in the design transformation process. It takes the RTL netlist as input and identifies the nets in the design that have an  high switching activity. \signed then uses specially crafted test vectors that toggle these selected nets, also termed as \emph{sensitive nets}, and records their response. The test vectors and response are stored in a challenge-response database. During the authentication phase, the IP vendor uses the challenge vectors on the design in-order to verify the provenance of the IP, to detect malicious modifications, or to detect the presence of Trojans. Unlike other watermarking schemes, even if the Trojan interferes with the watermark, we are still able to prove that the original circuit has been modified because we use more than one challenge vector in the authentication phase. Figure~\ref{fig:overallflow}a shows the watermark insertion process and Figure~\ref{fig:overallflow}b highlights the authentication process. 

\noindent The major contributions of the paper can be summarized as follows:
\textbf{\flushleft{1.}} It presents a novel watermarking technique termed as~\signed for digital IP blocks that uses a challenge-response protocol based interrogation scheme and can be applied to the IP at any stage in the design transformation process for authentication and provenance analysis. The watermarks inserted by ~\signed are design transformation invariant, i.e., they remain intact when a design goes through transformation from register transfer level (RTL) to gate level to transistor and layout. 
\textbf{\flushleft{2.}} We demonstrate that our proposed watermarking is lightweight, exhibits high structural coverage by showing that it can detect changes made to any portion of the circuit with high probability.
\textbf{\flushleft{3.}}We leverage the high coverage exhibited by~\signed to show that we can also detect malicious alterations in the design or Trojan attacks at an untrusted facility (e.g., foundry). To the best of our knowledge, this is the first watermarking technique that is capable of simultaneously detecting Trojan attacks.
\textbf{\flushleft{4.}}We also present possible integration of the proposed approach with any existing logic locking solution. While watermarking provides passive protection against IP piracy and helps with provenance analysis, logic locking provides active protection against piracy, reverse engineering and extraction of design secrets. By combining these two approaches, we achieve the benefits of both solutions towards comprehensive protection of hardware IPs, while also reducing overhead and strengthening each technique.
\textbf{\flushleft{5.}}We have developed a prototype CAD tool for the ~\signed framework. We use the CAD tool to evaluate the effectiveness of ~\signed  on ITC and ISCAS benchmarks. We show that the quality of the proposed watermarking, the coverage values of the internal nodes for Trojan detection, and hardware overheads obtained for these benchmark designs are highly promising.

\noindent The rest of this manuscript is organized as follows: Section 2 describes the ~\signed framework in detail. Section 3 presents our experimental set-up and evaluation results. Section 4 describes previous related work. Finally, we conclude the paper in Section 5.

\begin{figure*}[ht!]%
    \centering
    \subfloat[\centering Netlist before \signed watermark insertion. The nets highlighted in red are the sensitive nets]{{\includegraphics[width=.45\linewidth]{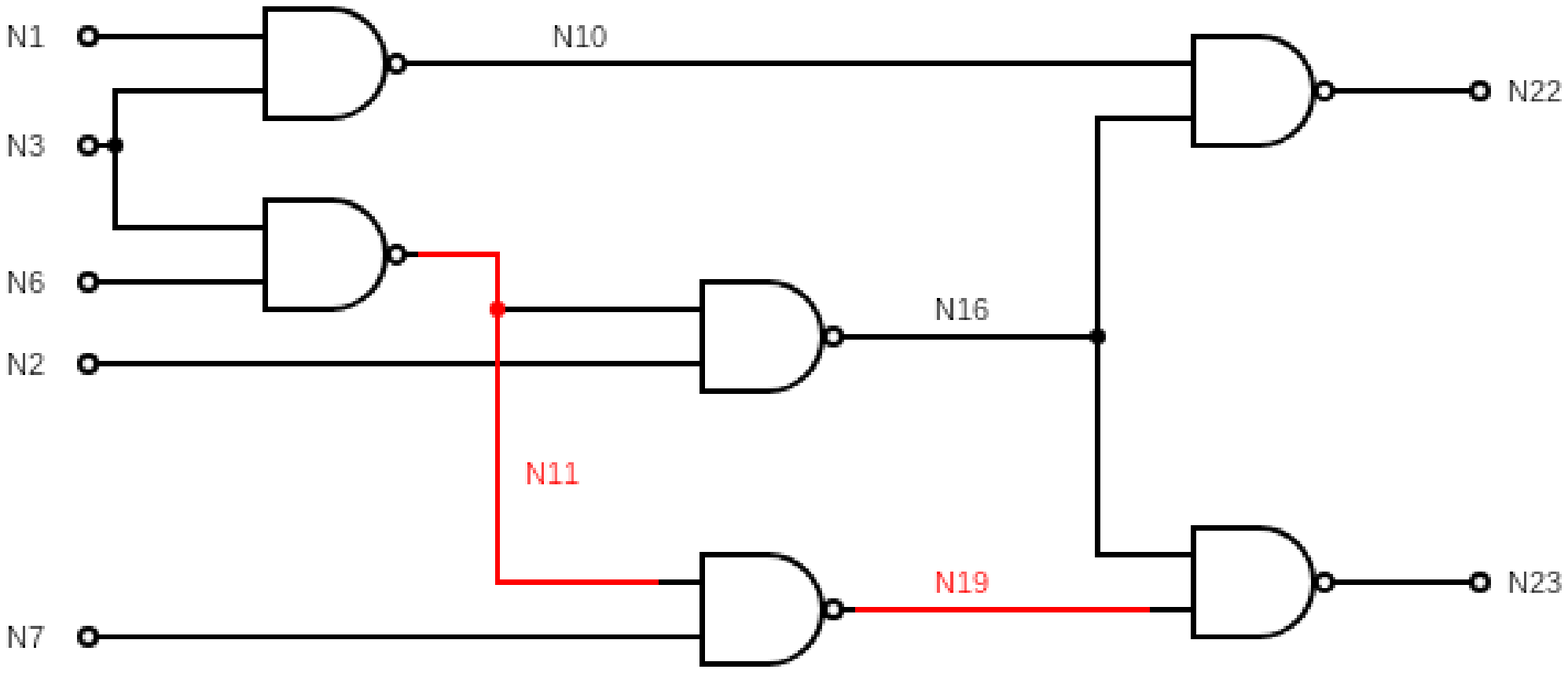} }}%
    \qquad
    \subfloat[\centering B. Watermark inserted netlist]{{\includegraphics[width=.34\linewidth]{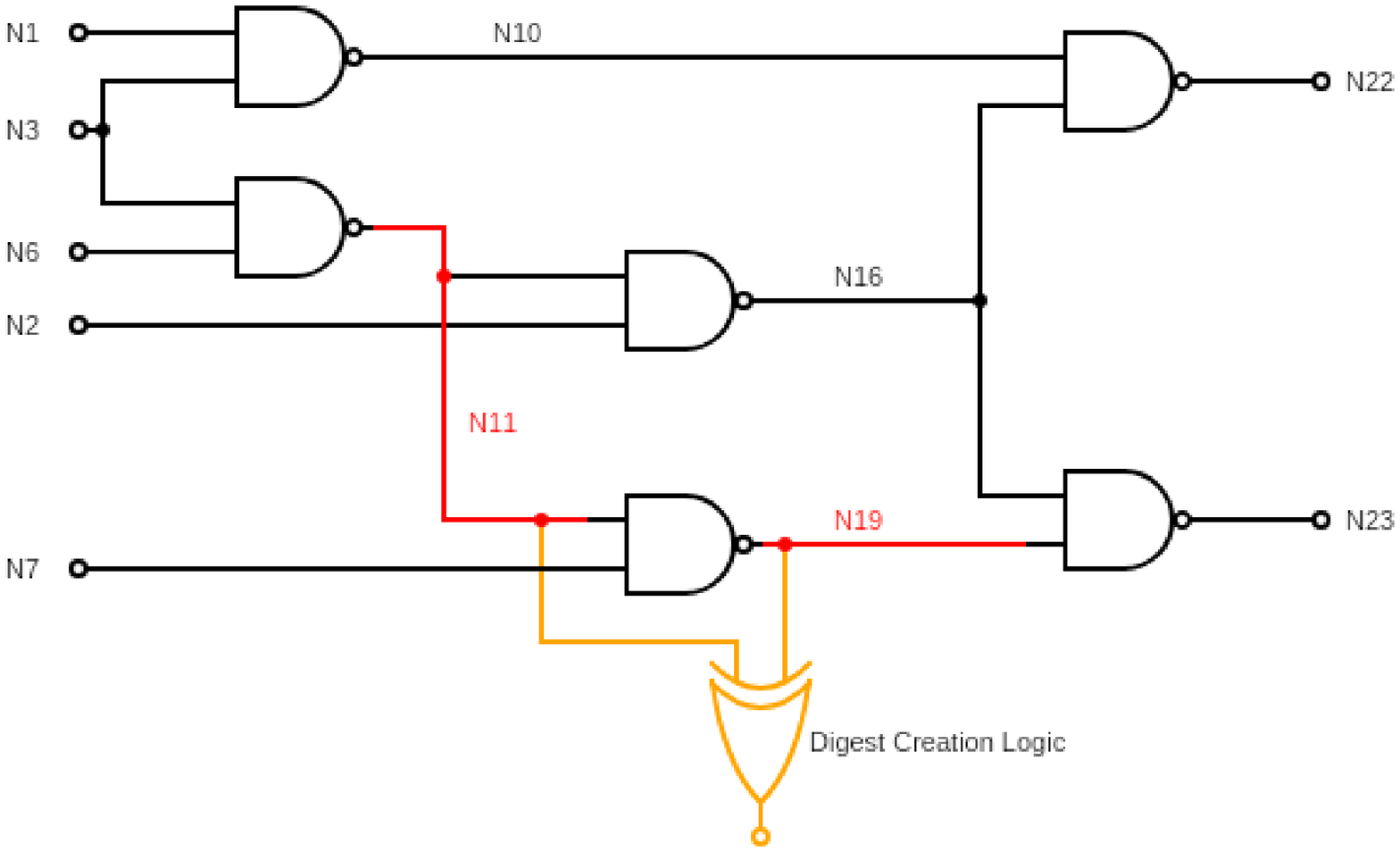} }}%
    \caption{Figure showing the various steps in the \signed process. For a given design, \signed identifies the \emph{sensitive nets} represented here in red and use these nets to generate a digest that is capable of both establishing provenance as well as detecting Trojans or other malicious modifications}%
    \label{fig:watermark}%
    \vspace{-10pt}
\end{figure*}
\begin{algorithm}[t!]

\caption{Sensitive Net Selection }
\scriptsize
\label{algo1}
  \SetKwInOut{Input}{Input}
  \SetKwInOut{Output}{Output}
  \Input{Circuit (C) represented as a Directed Graph}
  \Output{Set of sensitive nets (S)}
 $\mathbf{R_{C}}\leftarrow
 set~of~Random~Test~vectors~for~circuit~C$\\
 $S \leftarrow \varnothing $ \\ 
 Initialize circuit to a default value \\
 \ForEach{vector $R_{C}^{i}$ $\in$ $\mathbf{R}$}{ apply $R_{C}^{i}$ on C \\ 
 \ForEach{edge $e_{C}^{i}$ $\in$ C}{ store the value of $e_{C}^{i}$ for $R_{C}^{i}$}
 }
 
$T_{C}^{R}$ $\leftarrow$ $~Transition~matrix~with~e_{C}~columns~and~R_{C}~rows$\\

$D_{ij}$ $\leftarrow$ EuclideanDistance($T_{C}^{R}[i]$, $T_{C}^{R}[j]$)\\
Group nearby nets into a cluster $K_{C}^{i}$\\
$K_{C} \leftarrow $ Nets partitioned into $\mathbf{K}$ clusters based on $D_{ij}$\\

\ForEach{ partition $K_{C}^{i}$ $\in$ $K_{C}$ }
{
Sort the nets ($n_{i}$ ) in $K_{C}^{i}$ based on parameter $p_{n_{ij}}$ \\ 
where  $p_{n_{ij}}$ = $0.5*SW(n_{ij}) + 0.5*fanin(n_{ij})$

$e_{C}$ = random($K_{C}^{i}$, threshold)\\
S $\leftarrow$ S $\cup e_{C}$ \\
}

  
  
  
  
  
  \textbf{return} S
\end{algorithm}

\section{The \signed Framework \label{sec:framework}}
\label{sec:overview}
\noindent \signed consists of two major steps - watermark insertion and design authentication. Watermark authentication is performed on the RTL netlist and happens in the design house.On the other hand, design authentication is performed in the field in order to verify ownership and authenticity of the design. As shown in Figure~\ref{fig:overallflow}, the watermark insertion process can be split up into three separate components, namely sensitive net selection, digest logic insertion and challenge vector generation. We will now explain each of these with the help of Algorithm \ref{algo1} and Figure~\ref{fig:watermark}.

\noindent \textbf{Sensitive Net Selection:} In order to identify the sensitive nets within the circuit, we first simulate the design with a large number of random test vectors (lines 3-9 in Algorithm \ref{algo1}). The transitions of the Boolean values of the nets in the design are stored in the Transition matrix $T^{C}_{R}$. Next, we apply unsupervised learning techniques (K-means - lines 10 -13) to cluster the nets of the design based on their transitions in the Transition matrix. Nets that transition together i.e. undergo a change in their Boolean values in the same cycle are more likely to be in the same cluster. Once the $K_{C}$ clusters of the Transition matrix have been formed, we select one representative net from each cluster based on the weighted combination of the switching activity and fan-in values of the nets (lines 15 - 18). In Figure~\ref{fig:watermark}a, the nets in the circuit are split up into two clusters and nets N11 and N19 are selected as the representative nets. The set of all representative nets together are called the sensitive nets.

\noindent \textbf{Digest Creation Logic:} After having selected the sensitive nets from the design (N11 and N19 in Figure~\ref{fig:watermark}b), we insert a compaction logic into the design. This compaction logic reads the Boolean values of the sensitive nets and compresses it to create a digest. In Figure \ref{fig:watermark}b, the XOR gate (golden color) serves at the compaction logic. It reads the values of the two sensitive nets and creates a 1-bit digest.

\noindent \textbf{Challenge Vector Generation:} \signed relies on detecting tampering within the circuit by observing a mismatch between digest values in the tampered circuit and the un-tampered (golden) circuit. Therefore, we need to apply a set of specially chosen test vectors (called challenge vectors) and compare the digest values between the golden circuit and the circuit-under-test. These challenge vectors are chosen so as to maximize the probability of detecting any tampering, and therefore should maximize the probability of the sensitive nets undergoing a transition. In order to choose such vectors, we scan the previously generated set of random test vectors (line 1 in Algorithm \ref{algo1}) and identify those test vectors that cause the maximum number of sensitive nets to toggle. These challenge vectors are applied to the design and the digest values observed are stored in a database.

\noindent The second step of \signed~ i.e. design authentication, takes place in the field. the user verifies the provenance of the design by comparing the digest output with the digest output of the golden (fault-free) design when challenge vectors are applied.

\section{Evaluation}
\label{sec:results}

\noindent An effective watermark must posses the following properties: {\bf (1)} {\em immutability} - The various design transformations in the EDA flow must not affect the watermark; {\bf (2)} {\em undetectability} - The watermark should not be easily removable by an adversary, for e.g. an untrusted foundry; {\bf (3)} {\em uniqueness} - An adversary should not be able to clone the watermark and insert it into an arbitrary IP; {\bf (4)} {\em verifiability} - A legitimate user should be able to easily verify the authenticity of the design; {\bf (5)} {\em coverage} - Even minute modifications (or Trojans) in the design should be captured by the watermark; {\bf (6)} {\em cost} - The overheads introduced by the watermark should be acceptable for the use-case of the design. We now present the results of \signed on the ISCAS and ITC benchmarks and evaluate our watermarking scheme with reference to the above criteria.

\begin{figure}[!t]
  \centering
  \includegraphics[width=1.0\linewidth]{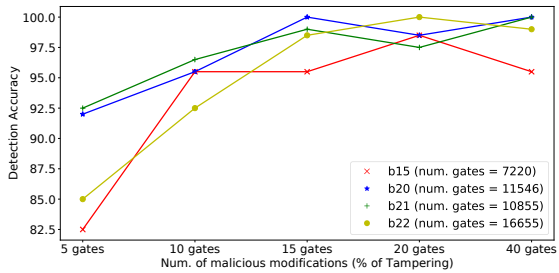}  
  \caption{Detection Accuracy vs Number of Modified Gates}
  \label{fig:davsfs}
  \vspace{-15pt}
\end{figure}

\noindent \signed~is capable of detecting malicious modifications in circuits even when the number of gates modified is less than 0.1\% of the total gate count. Figure \ref{fig:davsfs} shows the performance of \signed on a subset of the ITC benchmark. We observe that even when the percentage of gates modified is below 0.1\% of the total gate count, \signed is able to detect the modifications more than 80\% of the time. We would also like to highlight that when the percentage of gates modified is increased to 0.5\%, \signed is able to detect the modification every single time. 

\begin{figure}[!t]
  \centering
  \includegraphics[width=1.0\linewidth]{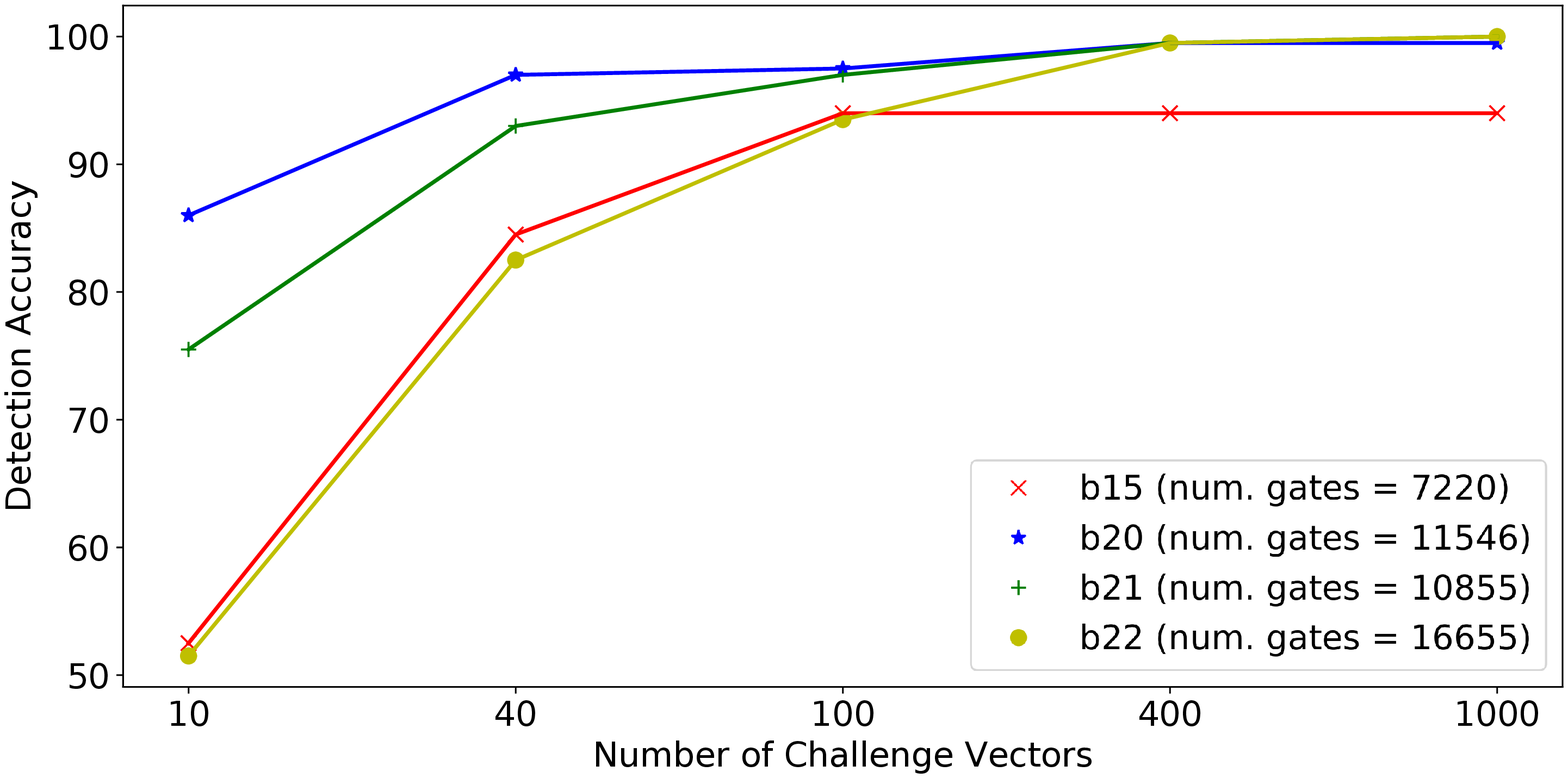}  
  \caption{Detection Accuracy vs Num. of Challenge vectors}
  \label{fig:davscr}
   \vspace{-15pt}
\end{figure}

\noindent The detection accuracy of \signed can be tuned using two parameters (knobs) - i) Size of the digest and ii) Number of challenge response pairs. Increasing the size of the digest allows us to capture more information about the circuit and therefore detect a greater number of malicious modifications or Trojans. Similarly, increasing the number of challenge vectors increases the probability of a modification or Trojan causing a mismatch in the Boolean values of the sensitive nets. Therefore, both of these parameters are directly proportional to the detection accuracy. Figure \ref{fig:davscr} clearly shows the dependence of the detection accuracy on the number of challenge vectors applied.

\begin{figure}[!t]
  \centering
  \includegraphics[width=1.0\linewidth]{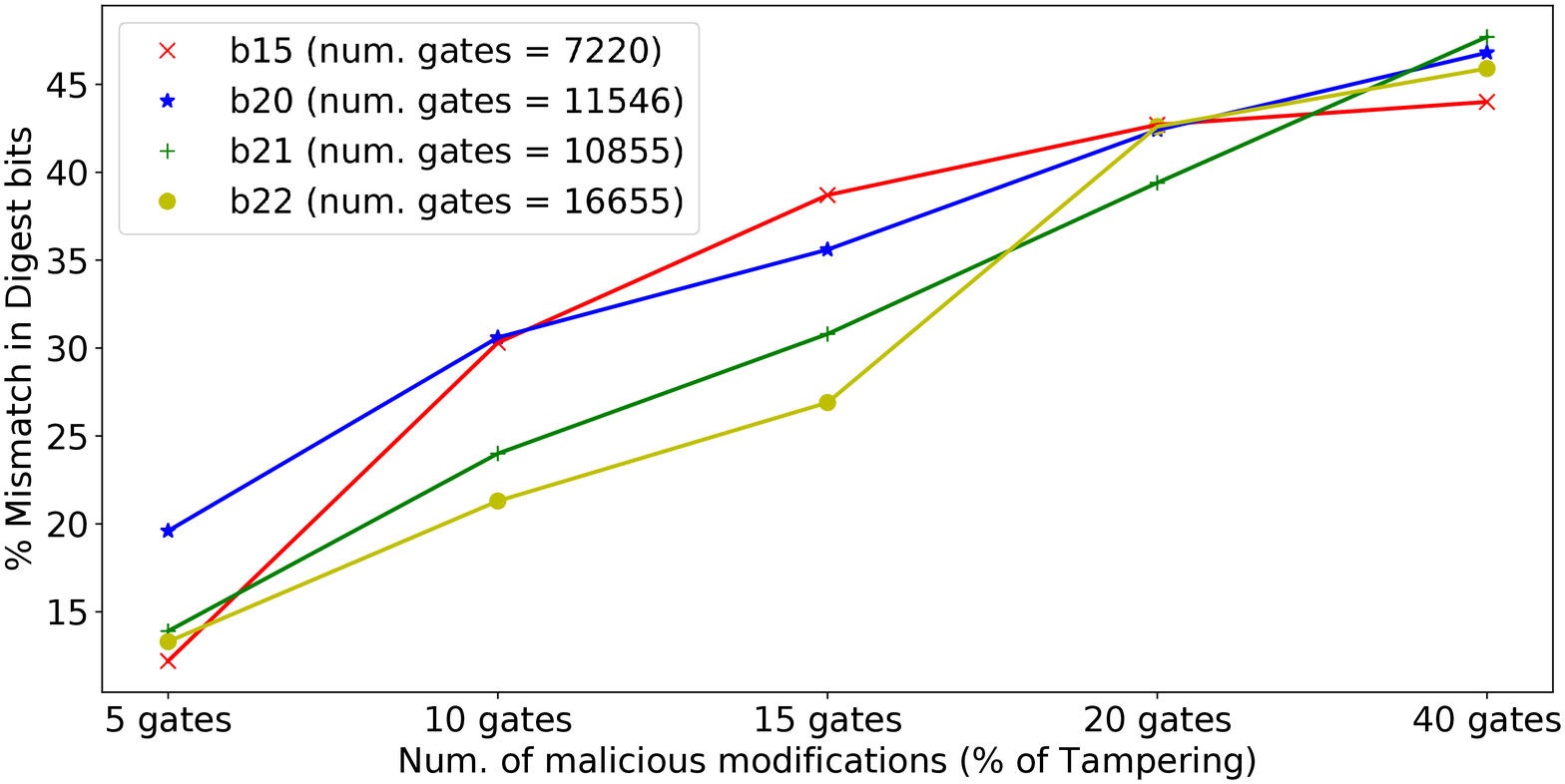}  
  \caption{Percentage change in Digest bits vs Num. of Modifications}
  \label{fig:sigvsfs}
   \vspace{-12pt}
\end{figure}

\noindent As discussed in Section~\ref{sec:overview}, \signed has two major steps viz, watermark insertion and watermark verification. Watermark verification is performed in-field, and under resource-constrained scenarios. Due to the sensitive and time critical nature of this process, it would be beneficial for the designer if they were provided with information regarding the nature of the malicious modification in the design. Figure~\ref{fig:sigvsfs} shows the relation between the size of the malicious modification and the number of mismatched bits in the digest. We observe that the degree of mismatch in digest bits is a reliable indicator of the number of gates that have been modified. For example, we observe that a modification as low as 5-bits produces a mismatch on one digest output for a 4-bit digest size. Thus, the designer can rely on the degree of change in the digest outputs to estimate the amount of modification. 

\noindent The results above clearly show that \signed satisfies the coverage criteria mentioned at the beginning of the section. It should also be noted that since \signed relies upon the stored golden responses to authenticate a circuit, it cannot be inserted into an arbitrary IP by an adversary (uniqueness property). Also, since the watermark does not change the functionality or usage of the design, the verification process is completely transparent to the end-user (easy verifiability).
\vspace{-10pt}
\subsection{Overhead analysis}
\noindent \signed modifies the design by injecting additional circuitry in-order to enhance the security of the IP. However, a significant overhead in terms of delay or area would render it unsuitable for low-resource domains. Hence, it is important to quantify the overheads of our proposed technique. We evaluate the impact of \signed-based watermarking on ISCAS85 and ITC benchmarks. The results for the largest benchmarks (>1000 gates) are presented in~\ref{tab:overhead}. 

\noindent We observe that \signed has negligible impact on the overall area, power and delay of the circuits, meaning that it also successfully satisfies the cost criteria.

\subsection{Security Analysis}
\noindent Although \signed is able to overcome tampering and Trojan insertion, it is still susceptible to removal attacks. In order to overcome this, we can combine our watermark insertion scheme with logic locking and consequently prevent reverse-engineering and removal attacks. \signed and logic locking will complement and strengthen each other. 

\begin{table}[!t]
\scriptsize
\caption{Table showing overheads for different benchmarks after 4-bit watermark insertion}
\begin{tabular}{|p{1cm}|p{0.5cm}|p{0.6cm}|p{0.4cm}|l|l|l|l|l|}
\hline
\multicolumn{1}{|p{0.9cm}|}{{\textbf{Benchmark}}} & \multicolumn{4}{c|}{\textbf{Before \signed~
}} & \multicolumn{4}{c|}{\textbf{After \signed~}} \\ \cline{2-9} 
\multicolumn{1}{|c|}{} & \begin{tabular}[c]{@{}l@{}}Gate\\ Count\end{tabular} & Area & Power & Delay & \begin{tabular}[c]{@{}l@{}}Gate\\ Count\end{tabular} & Area & Power & Delay \\ \hline
 c5315 & 1176 & 55.36  & 23.3 & 3.23 & 1204 & 57.853 &25.7 &5.29  \\ \hline
 c7552 & 1429 & 69.04 &  32.1 & 5.63 & 1457 & 71.53 & 35.0&5.66  \\ \hline
 b15 & 7220 &  394.86 & 6.2  & 8.53 & 7248 & 397.35 &6.2 &8.55   \\ \hline
 b17 & 22451 & 1226.0 & 14.0  &  10.24& 22479 & 1227.28  &14.0 &10.24  \\ \hline
 b18 & 48322 & 2641.0 & 33.4 &  10.09& 48350 & 2642.24  &33.4 &10.09  \\ \hline
 b19 & 90157 & 4953.7 & 35.8 &  11.84& 90185 & 4955.00  & 35.9& 11.84  \\ \hline
 b20 & 11546 & 618.90 & 10.8 &  8.96& 11574 & 621.39 & 10.8  &9.49 \\ \hline
 b21 & 10855 & 587.43 & 12.3 &  11.29& 10883 & 589.91 & 12.4 &11.61  \\ \hline
 b22 & 16655 & 887.17 &14.4  &  12.33& 16683 & 889.66 & 14.4&12.58 \\ \hline
\end{tabular}%
\label{tab:overhead}
\vspace{-10pt}
\end{table}

\section{Related Work}
\begin{table}[!t]
\scriptsize
\caption{Table comparing the different state-of-the-art watermarking schemes based on the level at which they are implemented, detectability, Transparency (Change in functional behaviour), and Coverage}
\begin{tabular}{|p{0.4cm}|p{1cm}|p{0.8cm}|p{0.6cm}|p{1cm}|p{0.7cm}|p{0.8cm}|}
\hline
\textbf{Work}& \textbf{Level}  & Type & Security & Transparency  & Coverage  & Overheads \\ \hline
 ~\cite{cuiconstraint} & Behavioural & Constraint& Medium & High & Medium & Medium \\ \hline
 ~\cite{quconstraint} & Physical & Constraint & Medium &  High&  Medium& High \\ \hline
 ~\cite{kahng} & Physical & Constraint& Medium &High  & Small & High  \\ \hline
 ~\cite{constraint} & Architectural &Constraint&  Medium&  High & Medium & Medium \\ \hline
 ~\cite{surkolay} &  Physical &FSM& Low & High & Small & High  \\ \hline
 
 \textbf{Ours} & Architectural &CRP& High & High & High  & Low  \\ \hline
\end{tabular}%
\label{tab:comparison}
\vspace{-10pt}
\end{table}

Existing IP watermarking techniques can be categorized into two distinct types based on the medium in which the watermark is embedded~\cite{survey1,survey2}. Constraint-based watermarking methods~\cite{survey1,cuiconstraint,quconstraint,constraint} or FSM-based watermarking techniques~\cite{robust,stateproperty,stateencoding}.  In constraint-based watermarking, the IP owner first creates a message that proves his ownership over the design. This message is then transformed into constraints for one specific phase of the IC design flow~\cite{constraint} and supplied to the EDA tool. The resulting design therefore carries the watermark through these constraints. On the other hand, in FSM based watermarking the watermark is embedded in the unused transitions of the state transition graph. Traversing the embedded sequence allows the designer to prove ownership of the design. However, these methods fail to meet the aforementioned properties of a robust watermarking technique. In particular, they cannot be easily verified by the consumer and can often be easily removed or circumvented by malicious actors. Moreover, they are unable to detect hardware Trojans or malicious modifications in the design. Therefore, there is a need for a robust watermarking scheme that detects tampering and meets the criteria delineated above. As we have shown in the previous sections, \signed satisfies all the properties needed for a robust watermark and is also capable of detecting hardware Trojans. We provide a qualitative comparison of \signed with other recent watermarking schemes in Table~\ref{tab:comparison}.  

\section{Conclusion}
\label{sec:conclusion}

IP Watermarking has emerged as a promising candidate for protection of semiconductor Intellectual Properties against a range of security threats. In this paper, we have introduced a novel challenge-response based IP watermarking scheme that is sensitive to extremely small modifications in the design. We insert the watermark by analyzing the structural properties of the design and achieve a high detection accuracy across multiple benchmarks. We also qualitatively compare \signed~with other recent watermarking approaches and evaluate it for the area, power and timing overheads. We show that such a scheme not only provides robust IP authentication and provenance analysis capability, but it also entails reliable detection of malicious alterations or Trojan attacks.

\bibliographystyle{ACM-Reference-Format} \bibliography{main}

\end{document}